\documentclass[pra, 12pt, onecolumn,showpacs,superscriptaddress]{revtex4}

\usepackage{graphicx}
\usepackage{color}

\begin{document}

\title{Entanglement and exotic superfluidity in spin-imbalanced lattices}

\author{V. V. Fran\c{c}a}
\affiliation{Institute of Chemistry, S\~{a}o Paulo State University, Brazil}

\date{\today}

\begin{abstract}
We investigate the properties of entanglement in one-dimensional fermionic lattices at the Fulde-Ferrell-Larkin-Ovchinnikov (FFLO) superfluid regime. By analyzing occupation probabilities, which are concepts closely related to FFLO and entanglement, we obtain approximate analytical expressions for the spin-flip processes at the FFLO regime. We also apply density matrix renormalization group calculations to obtain the exact ground-state entanglement of the system in superfluid and non-superfluid regimes. Our results reveal a breaking pairs avalanche appearing precisely at the FFLO-normal phase transition. We find that entanglement is non-monotonic in superfluid regimes, feature that could be used as a signature of exotic superfluidity.
\end{abstract}


\maketitle

\section{Introduction}

Entanglement has established its territory as one of the important resources for quantum information processes. More recently entanglement has been also considered as a probe for critical phenomena, such as quantum phase transitions \cite{fazio, amico, canovi, stas}, attracting therefore much interest in the condensed-matter community. While the investigation of entanglement in many-body interacting systems is increasingly desired, the realistic calculation of entanglement in such systems is however computationally too demanding in most of the cases. 

In this context the Hubbard model appears as an interesting approach, since it accounts, approximately, for spin and charge degrees of freedom of itinerant and interacting particles in a lattice, while still providing a proper description of nontrivial physics, such as superconductivity in solids, nanostructures and cold atoms in optical lattices. Moreover, entanglement measures for the Hubbard model have been investigated by several groups and nowadays there is a well defined measure for the ground-state entanglement of the Hubbard model \cite{zanardi, larsson}. For spatially homogeneous lattices there is in fact an analytical expression for entanglement \cite{28}, which may be used as input for density functional theory calculations in more realistic inhomogeneous systems \cite{27}.

The exotic superfluidity, the so-called Fulde-Ferrell-Larkin-Ovchinnikov (FFLO) phase \cite{1,2}, is one of the several interesting phenomena that have been investigated in solid systems within the Hubbard model. At low temperatures the FFLO state might emerge by the presence of external magnetic fields or by internal polarization as produced by spin-imbalanced populations. Although the exotic coexistence of superfluidity and magnetism has been investigated theoretical and experimentally since decades \cite{3} and state-of-the-art experiments with cold atoms \cite{15,16,17} and with organic superconductors \cite{organic} have been addressed this matter, there have been no unequivocal observations of FFLO superconductivity up to now.

From the theoretical point of view, thanks to powerful tools for describing complex many-body systems, such as quantum Monte Carlo, dynamical mean-field theory, density matrix renormalization group (DMRG) and density functional theory (DFT), substantial progress has been achieved in the FFLO business. Several FFLO features have been revealed, including spontaneous breaking of spatial symmetry in the pair correlation \cite{10, batrouni}, analytical expression for the critical polarization $P_C$ below which the FFLO superfluidity takes place \cite{fflo1}, regimes of stability \cite{stab} and temperature effects \cite{temp}. In contrast though, little is known about the relation between FFLO and entanglement, except from a few reports in gravity systems \cite{hol1, hol2, hol3}. 

Here we investigate the relation between entanglement and unconventional FFLO superfluidity in spin-imbalanced fermionic lattices described by the Hubbard model. We first analyze fundamental quantities to both entanglement and imbalance in solid systems $-$ the occupation probabilities $-$ and derive analytical approximations to describe spin-flip processes occurring at the FFLO regime. We thus calculate the exact ground-state entanglement entropy of the system via DMRG. Our results, supported by analytical and numerical calculations (DMRG and DFT), reveal a sudden breaking pairs phenomenon at the transition from the FFLO to the normal non-superfluid regime. We also find that entanglement is non-monotonic at the FFLO phase and, therefore, may be used as a witness of exotic superfluidity.

\section{Theoretical Model}

We consider one-dimensional superfluid lattices as described by the fermionic Hubbard model
\begin{equation}
H=-t\sum_{\langle ij\rangle\sigma}\hat{c}^\dagger_{i,\sigma}\hat{c}_{j,\sigma} +U\sum_i\hat{c}^\dagger_{i,\uparrow}\hat{c}_{i,\uparrow}\hat{c}^\dagger_{i,\downarrow}\hat{c}_{i,\downarrow},\label{hub}
\end{equation}
with attractive onsite interaction $U<0$ in units of the hopping parameter $t$, where $\hat{c}^\dagger_{i,\sigma}$ and $\hat{c}_{i,\sigma}$ are respectively creation and annihilation operators of particles with spin $\sigma=\uparrow, \downarrow$ at site $i$  and $L$ is the chain length. The imbalance is quantified by the polarization $P=(N_\uparrow-N_\downarrow)/N$, for a fixed number $N=N_\uparrow+N_\downarrow$ of particles. We adopt $N_\uparrow$ as the majority species, such that $P\geq 0$. 

The entanglement measure considered is the average single-site entanglement $S=\sum S_i/L$, defined as the average ground-state entanglement between a single site $i$ and the remaining chain sites $L-i$ \cite{ida}. Such bipartite pure systems have entanglement well quantified by the von Neumann entropy, which in the basis of occupation is written as  
\begin{eqnarray}
S_i&=&-w_{i,\uparrow}\log_2w_{i,\uparrow}-w_{i,\downarrow}\log_2w_{i,\downarrow}\\&&-w_{i,\uparrow\downarrow}\log_2w_{i,\uparrow\downarrow}-w_{i,0}\log_2w_{i,0},\nonumber\label{von}
\end{eqnarray}
where the occupation probabilities are \cite{28} 
\begin{eqnarray}
w_\uparrow&=&\frac{n}{2}(1+P)-w_{\uparrow\downarrow}\label{wup}\\
w_\downarrow&=&\frac{n}{2}(1-P)-w_{\uparrow\downarrow}\label{w2}\\
w_0&=&1-n+w_{\uparrow\downarrow}\\
w_{\uparrow\downarrow}&=&\frac{\partial e_0}{\partial U}.
\end{eqnarray}
Here $e_0\equiv e_0(n,P,U)$ is the ground-state energy per site and $(n, P, U)$ are local (onsite) quantities: particle density, polarization and interaction, respectively. The critical polarization $P_c$, which defines the transition from the FFLO phase to the partially polarized non-superfluid phase (normal phase), is obtained by solving the equality \cite{fflo1}
\begin{equation}
P_C(n,U)=\pm \left[\frac{4w_{\uparrow\downarrow}(n,P_C,U)}{n}-1\right].\label{pc}
\end{equation}

\section{Results and Discussion}

We start by analyzing one of the key quantities for both entanglement and imbalance: the unpaired majority probability $w_\uparrow$. For finite chains with a fixed number of particles, $P$ is enhanced via spin-flip processes, which increase the majority population by decreasing the minority one. There are two possible channels for such spin flips: either from unpaired minority states $|\downarrow\rangle$ (channel I) or from doubly occupied states $|\uparrow\downarrow\rangle$ (channel II). 

A first discussion about these channels at the FFLO regime was introduced in Ref. \cite{fflo1} through energy considerations. In this regime channel I is energetically the favored for spin flips, while channel II contribution is reduced due to the pairing mechanism. In spite of that, it was suggested that the polarization enhancement should be defined by a process combining both channels, I and II. This hypothesis was founded on the fact that $w_\downarrow$ is typically too small for $|U|>>t$ ($w_\downarrow \sim 0.05$ for $U=-8t$) to effectively sustain increasingly imbalances (up to the critical value, $P_c$). 

In order to {\it i)} probe the two channels' hypothesis, {\it ii)} quantify the contribution of each channel and {\it iii)} investigate whether the spin-flip processes also play a role at the precise FFLO-normal transition, we derive and analyze analytical approximate expressions for the spin-flip channels and compare them to numerical DMRG and DFT results. 

The spin-flip channel I, schematically $|\downarrow\rangle\rightarrow|\uparrow\rangle$, is modeled by assuming that only unpaired states contribute to produce $P$.  As a consequence, the doubly occupancy $w_{\uparrow\downarrow}(n,P,U)$ is considered as a constant in $P$. So given $n$ and $U$, one may use the doubly occupancy at the conventional BCS superfluid \cite{bcs}, i.e. for $P=0$: $w_{\uparrow\downarrow}\approx w_{\uparrow\downarrow}^{BCS}\equiv w_{\uparrow\downarrow} (P=0)$. Thus the ratio $w_\uparrow/w_{\uparrow\downarrow}$ within this model becomes linear with $P$ (from Eq.\ref{wup}),
\begin{eqnarray}
\text{\it channel I:}\hspace{0.6cm}
\frac{w_\uparrow}{w_{\uparrow\downarrow}}=\left(\frac{w_\uparrow}{w_{\uparrow\downarrow}}\right)^{BCS}+\frac{n}{2w_{\uparrow\downarrow}^{BCS}}P,
\end{eqnarray}
where both, the slope $n/(2w_{\uparrow\downarrow}^{BCS})$ and the offset
\begin{equation}
\left(\frac{w_\uparrow}{w_{\uparrow\downarrow}}\right)^{BCS}\equiv \frac{w_\uparrow (P=0)}{w_{\uparrow\downarrow}(P=0)}= \frac{n}{2w_{\uparrow\downarrow}^{BCS}}-1,\label{off}
\end{equation}
are obtained from the ground-state energy of the usual BCS superfluid.

In contrast, within the spin-flip channel II, schematically: $|\uparrow\downarrow\rangle\rightarrow|\uparrow\rangle,|\uparrow\rangle$, $P$ is assumed to enhance exclusively by spin flips of doubly occupied states and the rate at which $|\uparrow\rangle$ increases is twice the rate at which pairs decrease. So the linear expression for $w_\uparrow/w_{\uparrow\downarrow}$ with offset $(w_\uparrow/w_{\uparrow\downarrow})^{BCS}$ and slope $2$ is given by

\begin{eqnarray}
\text{\it channel II:}\hspace{0.8cm}
\frac{w_\uparrow}{w_{\uparrow\downarrow}}=\left(\frac{w_\uparrow}{w_{\uparrow\downarrow}}\right)^{BCS}+2P. 
\end{eqnarray}

Hence the combined linear model, channel I + II, has the same offset given by Eq.\ref{off} and slope given by $2+n/(2w_{\uparrow\downarrow}^{BCS})$, which simply sums the contribution of both channels. Therefore our model for the combined channel I + II is:

\begin{eqnarray}
\text{\it channel I+II:}\hspace{0.8cm}
\frac{w_\uparrow}{w_{\uparrow\downarrow}}=\left(\frac{w_\uparrow}{w_{\uparrow\downarrow}}\right)^{BCS}+\left(\frac{n}{2w_{\uparrow\downarrow}^{BCS}}+2\right)P.
\end{eqnarray}

\begin{figure}[t]
\centering  
 \includegraphics[width=14cm]{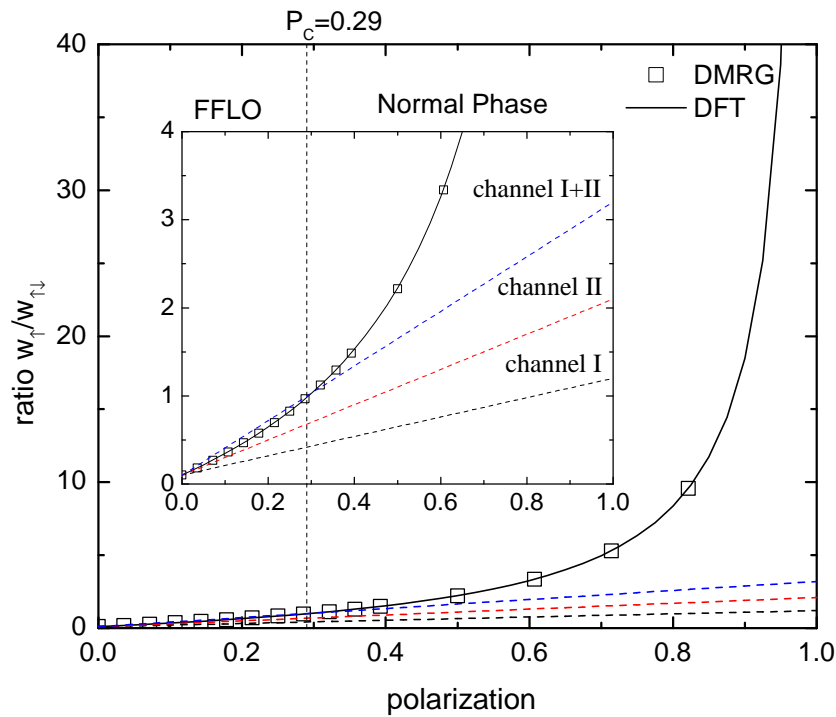}\vspace{-0.5cm}
    \caption{(Color online) Probabilities ratio $w_\uparrow/w_{\uparrow\downarrow}$ as a function of polarization $P$ as obtained by DFT-FVC \cite{fvc} and DMRG techniques for $L=80$, $n=0.7$ and $U=-8t$. Here $P_C\approx 0.29$ is indicated by a vertical dashed line. For $P<P_C$ (i.e. for $w_\uparrow/w_{\uparrow\downarrow}<1$) our combined channel I + II properly describes the pairing protection mechanism (see the inset), while for $P>P_C$ the ratio $w_\uparrow/w_{\uparrow\downarrow}$ deviates abruptly from the models, revealing a breaking pairs avalanche. }\label{channels}
\end{figure}

In Figure \ref{channels} we compare the performance of our analytical channels with numerical DMRG and DFT results. One verifies that channel I + II provides a very good description of the physics at the FFLO superfluid regime ($0<P\leq P_C$). In contrast, beyond $P_C$ the ratio diverges abruptly, becoming much greater than the predicted by the linear channel I + II. We interprete this {\it almost linearity} of the ratio with $P$ (for $P\leq P_C$) as a consequence of the pairing mechanism in the superfluid regime:  pairs are broken in a minimum rate necessary to enhance imbalance. On the other hand, for sufficiently strong polarizations pairs are broken in a rate much greater than the necessary to sustain the linear enhancement of $P$, what schematically corresponds to the process $|\uparrow\downarrow\rangle\rightarrow|\uparrow\rangle,|\downarrow\rangle$. Hence our results reveal that, beyond $P_C$, the spin-flip processes are responsible for a breaking pairs avalanche. As the phenomenon occurs precisely at $P=P_C$, we conclude that the spin-flip processes makes the FFLO phase fades away.


The effects of imbalance on the degree of entanglement in the system was also investigated, attempting to find a possible connection between FFLO and entanglement. We thus considered several situations at both FFLO and normal polarized regimes. To determine whether or not our system is in the FFLO state, we 
use the 1D unconfined phase diagram presented in \cite{fflo1}, obtained via DFT calculations \cite{fvc}. 

\begin{figure}[!h]
\centering  
\includegraphics[width=9.5cm]{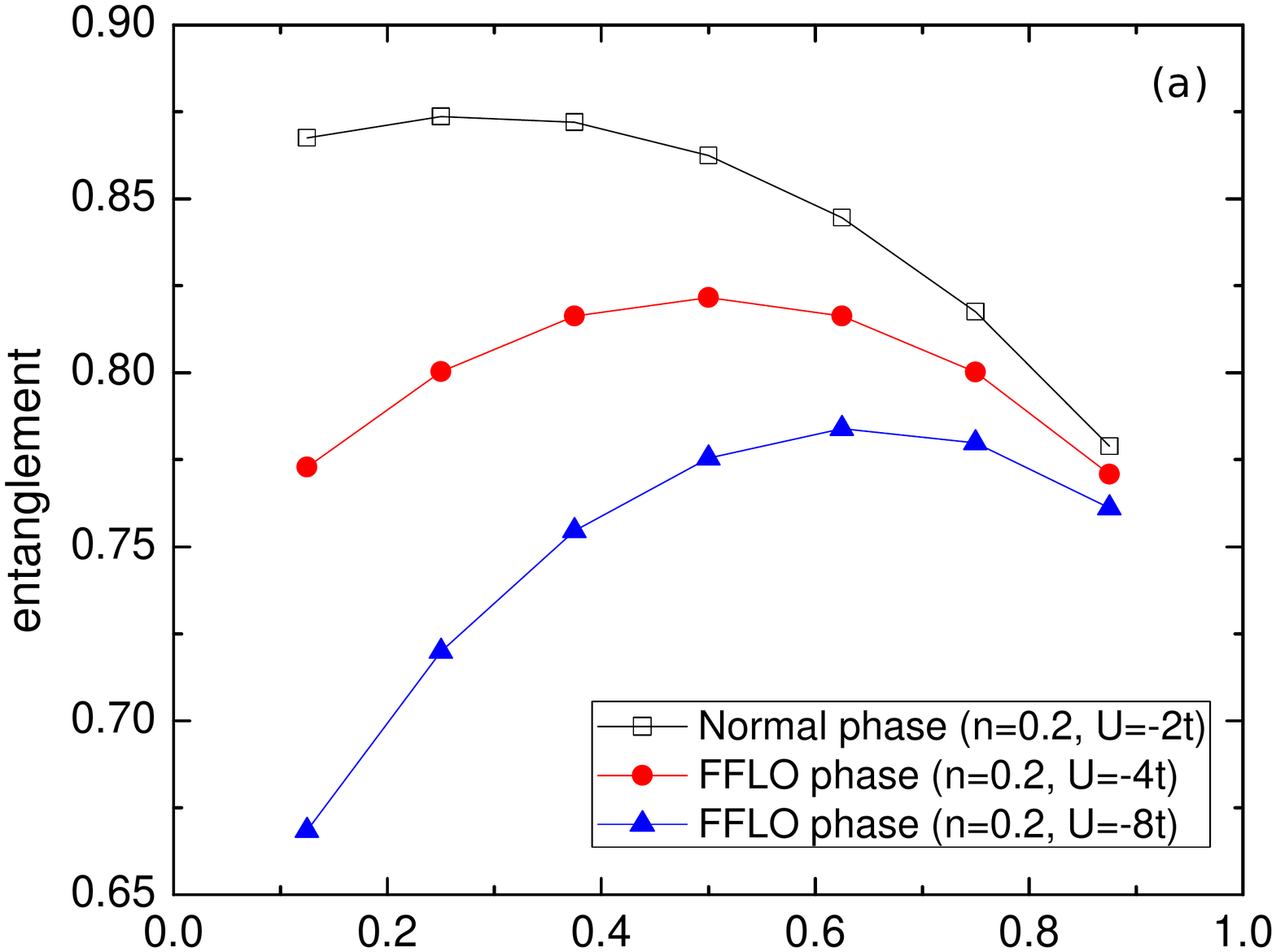}\vspace{-1.4cm}
    \includegraphics[width=9.5cm]{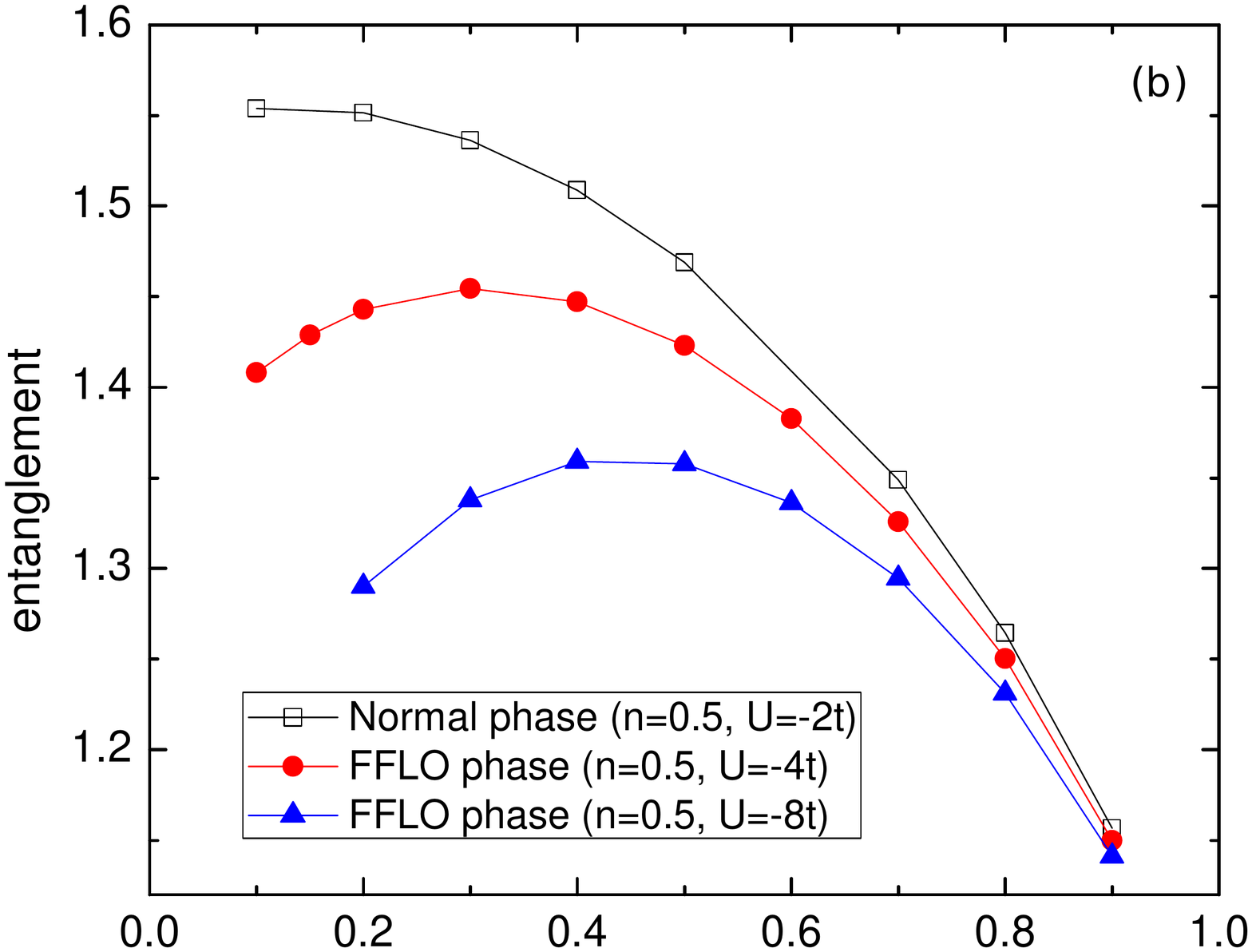}\vspace{-1.4cm}
      \includegraphics[width=9.5cm]{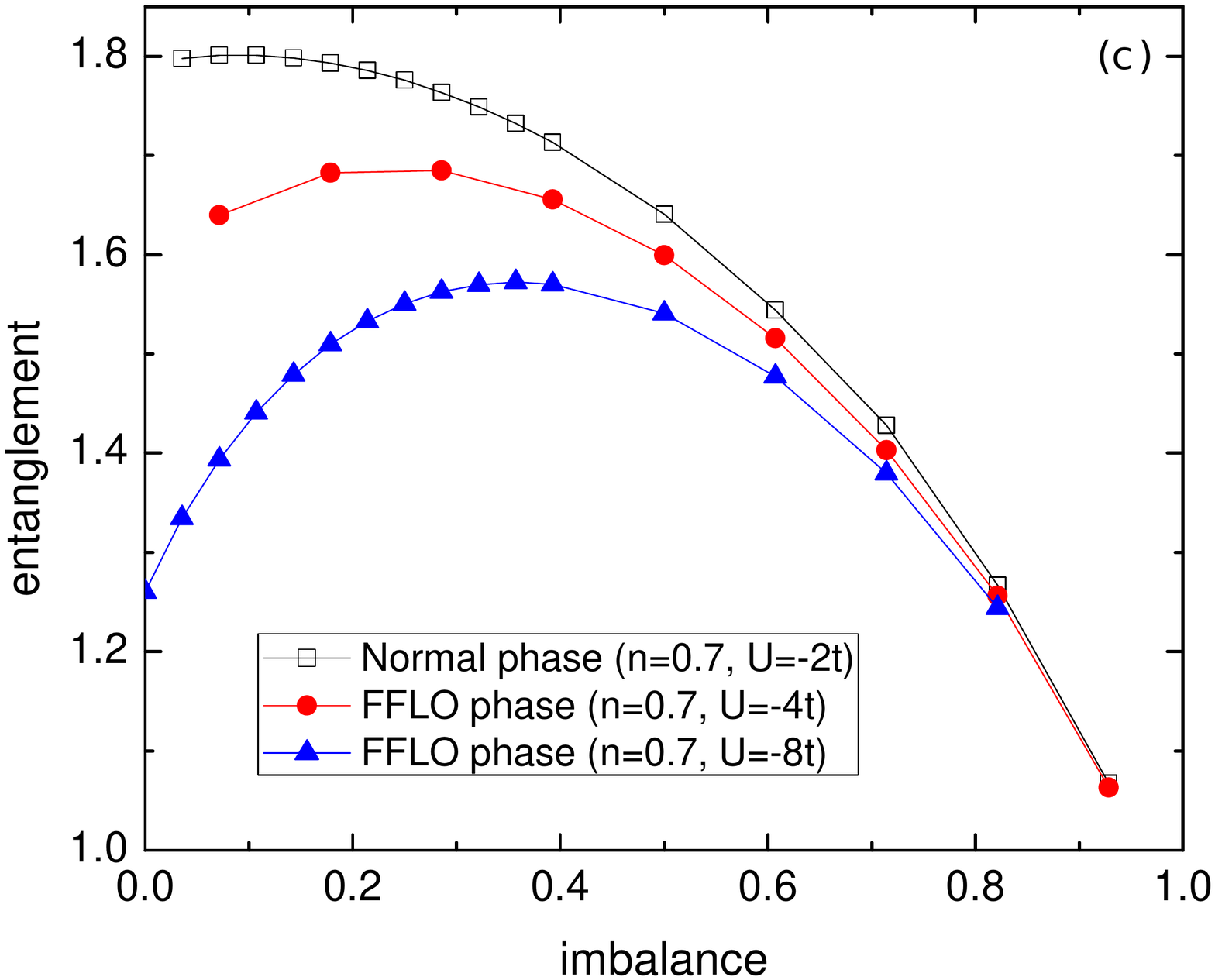}\vspace{-0.4cm}
    \caption{(Color online) Entanglement as a function of imbalance $P$ at the normal phase ($U=-2t$) and at the FFLO regime ($U=-4t$ and $U=-8t$) with filling factor (a) $n=0.2$, (b) $n=0.5$, and (c) $n=0.7$. In all cases entanglement is obtained via DMRG calculations for a chain of size $L=80$.} \label{ent}
\end{figure}

The entanglement entropy as a function of the imbalance for several choices of $n$ and $U$ is shown in Figure \ref{ent}. In general we find that the entanglement entropy is considerably sensitive to the imbalance. In particular, for a fixed $n$, we observe the decreasing of entanglement at the FFLO state in comparison to the normal phase. We also find that the entanglement at the FFLO phase depends on the interaction strength: the stronger the attractive interaction, the lower the degree of entanglement. We attribute this lower entanglement at the FFLO state and for larger $|U|$ to the considerable reduction of the degrees of freedom. It is caused by the pairing mechanism, which privileges $w_{\uparrow\downarrow}$ and $w_0$ and is even more robust for stronger attractive interactions. Similar reductions of the degree of freedom have been observed in repulsive systems \cite{28, 27}, but there it was induced by the particle density. Hence our findings are consistent with the idea that the entanglement entropy accounts for the degrees of freedom of the system and therefore plays the role of the order parameter in superconductors \cite{hol1}. 

We remark that the $P$ at which entanglement is maximum does not necessarily coincides with the critical polarization $P_C$ because $S^{max}$ occurs at the best balance among all the four degrees of freedom, while $P_C$ corresponds to the case $w_\uparrow=w_{\uparrow\downarrow}$, independently on the other two probabilities.

Finally, although the particular behavior of $S$ with $P$ depends on the specific values of $n$ and $U$, there is a general trend which allows one to distinguish the normal phase from the FFLO superfluid phase: $S$ is monotonic with $P$ at the normal phase, while it is non-monotonic at the FFLO state. Thus our findings not only show an explicit connection between entanglement and FFLO, but also indicate that entanglement might be a signature of the FFLO state. 

This non-monotonicity on entanglement has also been observed in homogeneous Hubbard chain in the BCS regime in which the polarization is due to an external magnetic field $h$ \cite{28}. For $U<0$ this system presents conventional superfluidity, described by BCS theory \cite{bcs}, while for $U\geq0$ the system is in a normal non-superfluid phase, described by Luttinger theory \cite{lutt}. Entanglement was found \cite{28} to be maximum at $U=0$ and $h=0$, for which the degrees of freedom are maximized, and to diminishe with $h$ monotonically for non-superfluid states ($U\geq0$) and {\it non-monotonically} for superfluid states ($U<0$). Hence we conclude that the non-monotonicity of entanglement as a function of internal ($P$) or external magnetic fields ($h$) may be used as a {\it general signature of superconductivity}. This also means that entanglement at the FFLO state of spin-imbalanced systems has a BCS-like behavior, similar to what has been reported in holographic superconductors \cite{hol2}.

\section{Conclusion}

In conclusion we have derived analytical expressions to describe approximately the spin-flip processes at the FFLO regime and have investigated the entanglement in several regimes: BCS, FFLO and normal non-superfluid phases. Our analytical approach was found to be in good agreement with numerical calculations obtained via DFT and DMRG. Our results show that FFLO state fades away due to the breaking pairs avalanche triggered by spin-flip processes. We have also found that entanglement entropy has specific behavior in each of the regimes, but that it is non-monotonic with internal and external magnetic fields at superfluid phases. We thus suggest that entanglement could be used as a witness of exotic superfluidity in fermionic systems. The effects of temperature remains to be investigated.

\section{Acknowledgments} This research was supported by the Brazilian agencies FAPESP (13/15982-3) and CNPq (448220/2014-8).

\end{document}